\documentclass[reqno,12pt]{amsart}

\usepackage{amsmath,amsthm,amssymb,color,graphicx}
\usepackage{amscd,verbatim}
\usepackage[all]{xy}

\theoremstyle{plain}
\newtheorem{theorem}{Theorem}

\theoremstyle{definition}

\theoremstyle{remark}
\newtheorem*{remark}{Remark}

\numberwithin{equation}{section}
\numberwithin{theorem}{section}
\numberwithin{figure}{section}
\numberwithin{table}{section}

\newcommand{\cE}{{\mathcal E}}

\newcommand{\cK}{{\mathcal K}}

\newcommand{\cL}{{\mathcal L}}

\newcommand{\CC}{{\mathbb C}}

\newcommand{\RR}{{\mathbb R}}
\newcommand{\TT}{{\mathbb T}}
\newcommand{\ZZ}{{\mathbb Z}}

\newcommand{\bZZ}{\ZZ}

\newcommand{\bfnu}{{\mathbf\nu}}
\newcommand{\bfom}{{\mathbf\omega}}

\newcommand{\txtfrac}[2]{{\textstyle{{\frac{#1}{#2}}}}}

\newcommand{\fA}{{\mathfrak A}}
\newcommand{\fg}{{\mathfrak g}}
\newcommand{\ft}{{\mathfrak t}}

\newcommand{\sfG}{{\mathsf G}}
\newcommand{\sfT}{{\mathsf T}}

\newcommand{\sfN}{{\mathsf N}}

\begin{document}

\title[T-duality for principal torus bundles and ...]{T-duality for principal torus 
bundles and dimensionally reduced Gysin sequences}

\author[P Bouwknegt]{Peter Bouwknegt}

\address[Peter Bouwknegt]{
Department of Physics and Mathematical Physics, 
and Department of Pure Mathematics \\
University of Adelaide \\
Adelaide, SA 5005 \\
Australia}
\email{pbouwkne@physics.adelaide.edu.au, 
       pbouwkne@maths.adelaide.edu.au}

\address[PB, after 1 Jan '05]{
Department of Theoretical Physics,
Research School of Physical Sciences and Engineering (RSPhysSE),
and Department of Mathematics,
Mathematical Sciences Institute (MSI),
Australian National University,
Canberra, ACT 0200, Australia}

\author[KC Hannabuss]{Keith Hannabuss}

\address[Keith Hannabuss]{
Mathematical Institute\\
24-29 St. Giles' \\
Oxford, OX1 3LB\\
England}
\email{kch@balliol.oxford.ac.uk}

\author[V Mathai]{Varghese Mathai}

\address[Varghese Mathai]{
Department of Pure Mathematics \\
University of Adelaide \\
Adelaide, SA 5005 \\
Australia}
\email{vmathai@maths.adelaide.edu.au}

\begin{abstract}
We reexamine the results on the global properties of T-duality 
for principal circle bundles in the context of a dimensionally reduced 
Gysin sequence.  We will then construct a Gysin sequence for principal 
torus bundles and examine the consequences.
In particular, we will argue that the T-dual of a principal torus bundle 
with nontrivial H-flux is, in general, a continuous field of noncommutative,
nonassociative tori.
\end{abstract}
\maketitle

\section{Introduction}

T-duality in String Theory, certainly from a local perspective, is an important 
and well-studied subject (see, e.g., \cite{Pol, Joh, Giv} and references therein for a comprehensive
review), but only recently have people begun to study the global properties of T-duality,
in particular in the presence of background fluxes.

In this paper we study global properties of T-duality for principal torus bundles 
in the background of NS H-flux.  We will do this by constructing a Gysin sequence
for principal torus bundles, which encodes the T-dual 
in terms of its invariants, e.g.\ a generalized Chern class.  
We will argue that in the most general case, the T-dual is
a bundle (or more precisely a continuous field) of noncommutative, 
nonassociative tori, generalizing all earlier partial results, \cite{BEMa, BEMb,
BHMa, BHMc, MRa, MRb}.  In a companion paper \cite{BHMc} we study 
the algebraic structures of the T-dual as arising from the results of this paper.
A similar conclusion, in the context of deformation quantization, was reached in
\cite{CS}.

The main observation in this paper is that
fluxes $[H] \in H^3(E)$ can be characterized by vector valued forms 
$(H_3, H_2, H_1, H_0)$ on the base manifold.  This is referred to as 
`dimensional reduction', or mathematically, as the Chern-Weil homomorphism.

T-duality for principal circle bundles  was treated geometrically in \cite{BEMa, BEMb, BS}, 
and dimensional considerations force the $H_0$ and  $H_1$
components of the H-flux to vanish in this case. The T-dual turns out to be another principal 
circle bundle with T-dual H-flux.
The arguments were extended in \cite{BHMa} to principal $\TT^n$-bundles with H-flux satisfying 
the 
condition that the $H_0$ and  $H_1$ components vanish. Then the T-dual turns out to be another
principal $\TT^n$-bundle with T-dual H-flux having vanishing $H_0$ and  $H_1$ components.
The analysis in \cite{MRa, MRb} shows that if one considers principal $\TT^n$-bundles with H-flux satisfying condition that just the $H_0$ component vanishes,
then one arrives at the surprising conclusion that the T-dual bundle has to 
have noncommutative tori as fibres, provided the $H_1$ component 
is non-zero. The weaker condition in \cite{MRa, MRb} permits non-vanishing $H_1$, but still
excludes non-zero $H_0$.
In this paper we will remove the last of these constraints and will allow for a
non-vanishing $H_0$ component.  
In this case, we arrive at the astonishing conclusion that 
the T-dual bundle has to have nonassociative tori as fibres, taking it even
beyond the normal range of non-commutative geometry.  The algebraic structure, 
generalizing that of \cite{MRa,MRb} is discussed in \cite{BHMc}.

The paper is organized as follows.  In order to explain our ideas as carefully as 
possible, we first recall in Section 2 how the Buscher rules
encode the global aspects of T-duality.  Then we review how those rules are related 
to the Gysin sequence for principal circle bundles.  In the latter part of Section 2 we
make closer contact with the usual form of the Buscher rules by dimensionally 
reducing the Gysin sequence.  We also discuss how T-duality gives rise to an
isomorphism on twisted cohomology in this dimensionally reduced setting.  Motivated 
by the discussion in Section 2, we generalize the dimensional reduction to (higher 
rank) principal torus bundles in Section 3.  We derive a Gysin sequence for principal 
torus bundles (which, to the best of our knowledge, has not appeared elsewhere). In Section 4
we  deduce from the Gysin sequence of Section 3
that the the T-dual of a principal torus bundle with H-flux is, in general,
a continuous field of noncommutatitve, nonassociative tori over the base.
We discuss both the T-duality group and a simple example in this setting.
We end, in Section 5, with some conclusions and open problems.
In an appendix we briefly discuss duality from the operator algebraic perspective, which might
be useful for the reader to relate the results of this paper to the results of, in particular,
\cite{MRa,MRb}, as well as the companion paper \cite{BHMc} where we explore the 
nonassociative structures arising in this paper in more detail.

\section{Principal circle bundles}

To motivate our construction, in the case of principal torus bundles, we
first look at the simplest case, name that of a principal circle bundle.

\subsection{A bit of notation}

Throughout this paper, we will denote by $\Omega^k(E)$, $Z^k(E)$, and $B^k(E)$,
the spaces of $k$-forms, closed $k$-forms, and exact $k$-forms on a smooth
manifold $E$, respectively.  The de-Rham cohomology of $E$ us defined as
$H_{\text{dR}}^k(E) = Z^k(E)/B^{k}(E)$.  The integer, or \v Cech, cohomology of $E$ will
be denoted by $H^k(E,\ZZ)$.  For clarity, we will restrict the 
presentation in this paper to de-Rham cohomology only, or rather to the image
$H^k(E) = i (H^k(E,\ZZ)) \subset H_{\text{dR}}^k(E) $ of integer cohomology in de-Rham
cohomology (i.e.\ forms with integral periods), but most of the results
of this paper will generalize to integer cohomology without too much effort.

\subsection{The Buscher rules revisited}
\label{secBB}

We start with a principal $S^1$-bundle $\pi : E \to M$, supported by an H-flux
$[H] \in H^3(E)$.  If $A$ denotes a connection 1-form on $E$, and $\bar g$ a metric
on the base $M$, then $E$ carries a canonical metric $g = \bar g + A\otimes A$.
If $\kappa$ denotes the Killing vector corresponding to the $S^1$-isometry, 
we may choose a (de-Rham) representative $H$ of $[H]$, satisfying the invariance 
condition $\cL_\kappa H =0$.  Again, locally $H=dB$ for a two-form $B$, and we 
will assume that $B$ can be chosen such that $\cL_\kappa B =0$ (this is not a necessary 
requirement, but it will slightly simplify the discussion below).

Locally, we can choose coordinates $x^M = (x^\mu,x^0) \equiv (x^\mu,\theta)$ on $E$ such that the
Killing vector of the $S^1$-isometry is given by $\kappa = \partial /\partial \theta$.  
The invariance conditions $\cL_\kappa H =\cL_\kappa B=0$ then simply imply
that the components $H_{LMN}$ and $B_{MN}$ do not depend on $\theta$.

Furthermore, locally we can choose the connection $A = A_M\, dx^M = 
d\theta + A_\mu dx^\mu$, where 
$A_\mu dx^\mu \in \Omega^1(M)$.  I.e.,
\begin{align}
g & = \bar g + A\otimes A = \bar g_{\mu\nu}\, dx^\mu \otimes dx^\nu + (d\theta +A_\mu dx^\mu)^2 
\,, \nonumber \\
B & = \txtfrac12 B_{\mu\nu} \, dx^\mu \wedge dx^\nu + B_\mu \,  dx^\mu \wedge (d\theta +
 A_\nu dx^\nu) \,,
\end{align}
where the components $A_\mu$, $B_{\mu\nu}$ and $B_\mu$ do not depend on $\theta$.
Physically, the decomposition of $g$ and $B$ above is referred to as dimensional reduction.
In terms of matrices, 
the metric and B-field components are
\begin{equation} g_{MN} = 
\begin{pmatrix}  \bar g_{\mu\nu}+A_\mu
A_\nu & A_\mu \\ A_\nu & 1 \end{pmatrix} \,,\qquad
B_{MN} = \begin{pmatrix} B_{\mu\nu}  +  (B_\mu A_\nu - 
A_\mu B_\nu )  & B_{\mu} \\ -B_{\nu} & 0 \end{pmatrix} \,.
\end{equation}

The Buscher rules \cite{Bus} 
\begin{align}
\widehat g_{\mu\nu} & =  g_{\mu\nu} - \frac{1}{g_{00}} 
\left( g_{\mu0}g_{\nu0} - B_{\mu0}B_{\nu0} \right) \,, & 
\widehat g_{0 0} & =  \frac{1}{g_{00} } \,, & 
\widehat g_{\mu0}  & =  \frac{B_{\mu0}}{g_{00}}\,, \nonumber \\
\widehat B_{\mu\nu} & =  B_{\mu\nu} - \frac{1}{g_{00}} 
\left( g_{\mu0}B_{\nu0} - g_{\nu0} B_{\mu0}\right) \,, & 
\widehat B_{\mu0} & =  \frac{g_{\mu0}}{g_{00}} \,,
\end{align}
give
\begin{equation} \label{eqAc}
\widehat g_{MN} = 
\begin{pmatrix}  \bar g_{\mu\nu}+B_\mu
B_\nu & B_\mu \\ B_\nu & 1 \end{pmatrix} \,,\qquad
\widehat B_{MN} = \begin{pmatrix} B_{\mu\nu} & A_{\mu} \\  -A_{\nu} & 0 \end{pmatrix} \,.
\end{equation}
I.e., with the choices above, T-duality locally simply corresponds to the interchange
$A_\mu \leftrightarrow B_\mu$.

Denoting the coordinate of the dual circle by $\hat \theta$, we can interpret
$\widehat A = d\hat\theta + B_\mu dx^\mu$, locally, as a connection on a dual 
circle bundle $\widehat \pi : \widehat E \to M$. 
We deduce from Eqn.~\eqref{eqAc} that on the correspondence space $E\times_M \widehat E=
\{ (x,\hat x) \in E\times \widehat E \ | \ \pi(x) = \widehat \pi(\hat x) \}$,
with local coordinates $(x^\mu,\theta,\hat\theta)$,
\begin{equation}
\widehat B = B  +   A \wedge \widehat A -  d\theta  \wedge d\widehat \theta\,,
\end{equation}
so that
\begin{equation} \label{eqAa}
\widehat H - H =  d(A\wedge \widehat A) =  F \wedge \widehat A - A \wedge \widehat F\,,
\end{equation}
where $F=dA$, and $\widehat F = d\widehat A$ are the curvatures of $A$ and $\widehat A$,
respectively, and are (globally) defined forms on $M$.
Equation \eqref{eqAa} actually makes sense globally on $E\times_M \widehat E$.
Rewriting this equation as
\begin{equation*}
H - \widehat F\wedge A = \widehat H - F \wedge \widehat A \,,
\end{equation*}
we see that the left hand side is a form on $E$, while the right hand side 
is a form on $\widehat E$.  Thus, in order to have equality, 
we conclude that both have to equal a form $H_3$ defined on $M$.  I.e.
\begin{align} \label{eqBAa}
H & = H_3 + \widehat F \wedge A \,, \nonumber \\ 
\widehat H & = H_3 + F \wedge \widehat A  \,.
\end{align}
We note that these equations imply that
\begin{equation} \label{eqAb}
F = \widehat\pi_* \widehat H \,,\qquad
\widehat F = \pi_* H\,,
\end{equation}
which is the statement that 
H-flux and first Chern class of the circle bundle are exchanged under T-duality
\cite{BEMa,BEMb}.
 
\subsection{Dimensionally reduced Gysin sequence}
\label{secBC}

Let us first review 
how the global content of the Buscher rules, i.e.\ Eqn.\ \eqref{eqAb},
is encoded in the Gysin sequence for the principal circle bundle $\pi:E\to M$ 
(cf.~\cite{BEMa,BEMb}).
Principal circle bundle are classified, up to isomorphism, by 
the Euler class $\chi(E)\in H^2(M,\ZZ)$, or equivalently, by the first Chern class $c_1(L_E)
\in H^2(M,\ZZ)$
of the associated line bundle $L_E = E \otimes_{U(1)} \CC$.  
Given a principal circle bundle $\pi:E\to M$, we have the pull-back map
$\pi^*: H^k(M) \to H^k(E)$, as well as the push-forward map (`integration over the $S^1$-fibre')
$\pi_* : H^k(E) \to H^{k-1}(M)$.  These maps nicely fit into a long exact sequence
in cohomology, the so-called Gysin sequence \cite{BT}.\footnote{This Gysin 
sequence also holds in integer cohomology,
but for simplicity of presentation we restrict ourselves to de-Rham cohomology throughout
the paper.}
\begin{equation} \label{eqBBa}
\begin{CD}
\ldots @>>> H^k(M) @>\pi^*>> H^k(E) @>\pi_*>> H^{k-1}(M) @> \delta >> 
H^{k+1}(M) @>>>  \ldots
\end{CD}
\end{equation}
where the map $\delta: H^{k-1}(M) \to H^{k+1}(M)$ is given, on a representative
$\omega$ of a class in $H^{k-1}(M)$, by $\delta\omega =  F\wedge \omega$.  Here,
$F$ is (a representative for) the Euler class of $E$, i.e.\ the curvature of 
a connection on $E$.

Considering the $k=3$ segment of the Gysin sequence \eqref{eqBBa}, we
see that any class $[H]\in H^3(E)$, i.e.\ any H-flux on $E$, gives rise to a class
$\pi_*[H] \in H^2(M)$, which can be interpreted as the equivalence class $[\widehat F]$ of the 
curvature $\widehat F$ of a T-dual circle bundle $\hat\pi : \widehat E \to M$.\footnote{The
isomorphism class of this T-dual bundle is only unique up to torsion, but would be unique
if we would have presented the analysis in integer cohomology.}
Furthermore, we have $[F]\wedge [\widehat F] \equiv 0$ in $H^4(M)$.
Conversely, by considering the Gysin sequence corresponding to the T-dual
circle bundle $\hat\pi : \widehat E \to M$, we conclude from $[F]\wedge [\widehat F] =
[\widehat F] \wedge [F] \equiv 0$, that $[F] = \hat \pi_*[\widehat H]$, for some class
$[\widehat H] \in H^3(\widehat E)$.  This is precisely the content of Eqn.\ \eqref{eqAb}.

{}From the Gysin sequence we can of course only determine 
the element $[\widehat H] \in H^3(\widehat E)$ up to an element in $\pi_*(H^3(M))$.  
To fix this ambiguity, we need some extra input.  The extra input, of course, is that 
T-duality should not affect that part of the H-flux that `lives' on the base manifold $M$.  
This is equivalent to demanding that $p^*[H] - \hat p^*[\widehat H] \equiv 0$ in
$H^3(E\times_M \widehat E)$ where $E\times_M \widehat E
=\{ (x,\hat x) \in E\times \widehat E \ | \ \pi(x) = \widehat \pi(\hat x) \}$ is the correspondence 
space, and the projections $p$ and $\hat p$ are defined in the following commutative diagram
\begin{equation*}
\xymatrix @=5pc @ur { E \ar[d]_{\pi} & 
E\times_M  \widehat E \ar[d]_{\hat p} \ar[l]^{p} \\ M & \widehat E\ar[l]^{\hat \pi}}
\end{equation*}

With a bit more work, using again the Gysin sequence, one can actually argue that at the 
level of forms, we reproduce Eqns.\ \eqref{eqAa}-\eqref{eqAb} (see \cite{BEMa} for details).

Obviously, there is a great deal of similarity between the analysis in Section \ref{secBB}  and the 
discussion from the point of view of the Gysin sequence above.  This similarity can be
further illuminated by `dimensionally reducing' the Gysin sequence \eqref{eqBBa}.  The resulting
dimensionally reduced Gysin sequence will immediately present to us how the above analysis
should be generalized to higher rank principal torus bundles.

To explain the ideas, let us start by considering  
the trivial principal circle bundle $\pi: M \times S^1 \to M$.
By the K\"unneth theorem we have 
\begin{equation} \label{eqBa}
H^k(E) \cong \bigoplus_{p+q=k} \left( H^p(M) \oplus H^q(S^1) \right)  
\cong H^k(M) \oplus H^{k-1}(M) \,.
\end{equation}
Explicitly, if we denote the generator of $H^1(S^1)$ by $d\theta$, normalized 
such that $\pi_*(d\theta) = \int d\theta =1$, then the 
isomorphism \eqref{eqBa} is given by 
\begin{equation}  \label{eqBb}
(\omega_k, \omega_{k-1}) \mapsto \omega_k +  d\theta \wedge \omega_{k-1} \,, 
\end{equation}
with inverse, for $\Omega\in \Omega^k(E)_{\text{inv}}$
\begin{equation} \label{eqBc}
\Omega \mapsto 
(\omega_k,\omega_{k-1}) \equiv (\Omega - d\theta \wedge \pi_*\Omega,  \pi_*\Omega)\,.
\end{equation}
Note that, in particular, $\pi_* ( \Omega - d\theta \wedge \pi_*\Omega)  = 0$, so that
$\omega_k$ can indeed be identified with an element of $\Omega^k(M)$.

Now consider a nontrivial circle bundle $\pi:E\to M$.   Choose a representative curvature 
$F\in \Omega^2(M)$ such that
$[F]=c_1(L_E)$, with connection $A\in \Omega^1(E)$, i.e.\ $\pi^*F =dA$, normalized 
such that $\pi_*A=1$.
The question arises whether we can still characterize classes in $H^k(E)$ by
forms living on the base manifold.  

Let $\Omega$ be a representative of an element in $H^k(E)$.  We would like to mimic 
\eqref{eqBc}, but of course, in the more general case, the element $d\theta$ is not a 
globally defined 1-form on $E$.  Instead we can make use of the connection $A$.
Again, let $\kappa$ denote the (globally defined) Killing vector field corresponding to
the $U(1)$-isometry, and let $\Omega^k(E)_{\text{inv}}$ denote the space of $k$-forms
invariant under the isometry, i.e.\ $\cL_\kappa \Omega=0$.  Then we have a map
\begin{equation} \label{eqBBi}
f_A : \Omega^k(M)\oplus \Omega^{k-1}(M) \to \Omega^k(E)_{\text{inv}} \,,\qquad
(\omega_k,\omega_{k-1} ) \mapsto \omega_{k} + A \wedge \omega_{k-1} \,.
\end{equation}
with inverse 
\begin{equation} \label{eqBBh}
f_A^{-1} : \Omega^k(E)_{\text{inv}} \to \Omega^k(M)\oplus \Omega^{k-1}(M)\,,\qquad
\Omega \mapsto (\Omega - A\wedge \pi_*\Omega, \pi_*\Omega) \,.
\end{equation}

A simple computation shows 
\begin{equation*}
(d\circ f_A)(\omega_k,\omega_{k-1}) = (d\omega_k + F\wedge \omega_{k-1} ) -
A \wedge d\omega_{k-1} \,,
\end{equation*}
thus, upon defining a modified differential
$D : \Omega^k(M)\oplus \Omega^{k-1}(M) \to
\Omega^{k+1}(M)\oplus \Omega^{k}(M)$,  by
\begin{equation}
D(\omega_k,\omega_{k-1} ) = (d\omega_k + F\wedge \omega_{k-1}, -d\omega_{k-1})\,,
\end{equation}
we have $d\circ f_A = f_A \circ D$. 
It is straightforward to check that $D^2=0$, and hence that $D$ defines a 
cohomology $H_D^k(M) \equiv H^k(\Omega^\bullet(M)\oplus \Omega^{\bullet -1}(M), D)$.
Furthermore, because of the commutativity of the diagram
\begin{equation*} \begin{CD}
\Omega^k(M)\oplus \Omega^{k-1}(M) @>\cong>f_A>  \Omega^k(E)_{\text{inv}}\\
@VDVV  @VVdV \\
\Omega^{k+1}(M)\oplus \Omega^{k}(M) @>\cong>f_A>  \Omega^{k+1}(E)_{\text{inv}} \end{CD}
\end{equation*}
we have the result 
\begin{equation} \label{eqBBj}
H^k(E) \cong H_D^k(M)\,.
\end{equation}
While the explicit isomorphism \eqref{eqBBi} depends on the choice of connection $A$, it
is easily verified that the isomorphism \eqref{eqBBj} is independent of the choice of $A$.

Now that we have a globally defined dimensional reduction of forms, Eqn.\ \eqref{eqBBh}, 
and an identification of cohomology, Eqn.\ \eqref{eqBBj}, it is straightforward to
dimensionally reduce the Gysin sequence \eqref{eqBBa}.  The result is the following 
exact sequence 
\begin{equation} \label{eqBg}
\begin{CD}
\ldots @>>> H^k(M) @>\pi^*>> H^k_D(M) @>\pi_*>> H^{k-1}(M) @>\delta>> 
H^{k+1}(M) @>>>  \ldots
\end{CD}
\end{equation}
where the various maps, on representatives of the cohomology, are given by
\begin{align*}
\pi^* : H^k(M) \to H^k_D(M)\,, \qquad& \pi^*(\omega_k) = (\omega_k,0) \,,\\
\pi_* : H^k_D(M) \to H^{k-1}(M)\,, \qquad & \pi_*(\omega_k,\omega_{k-1}) = \omega_{k-1} \,,\\
\delta : H^{k-1}(M) \to H^{k+1}(M)\,, \qquad & \delta(\omega_{k-1}) = F \wedge \omega_{k-1} \,.
\end{align*}
It is interesting to observe that we 
can consider the sequence \eqref{eqBg} in itself, without making reference to
any principal circle bundle, but just defined by a certain representative of $[F] \in H^2(M)$.

Usually, the proof that \eqref{eqBBa} defines a long exact sequence in
cohomology is quite involved.  The standard proof is given by examining the 
Leray spectral sequence corresponding to 
the principal circle bundle (see e.g.\ \cite{BT}).  After dimensional reduction, however,
the proof is remarkably simple and does not require sophisticated techniques.  To illustrate  
this point, we present the proof below.

\begin{theorem} 
The sequence \eqref{eqBg} defines an exact complex.
\end{theorem}

\begin{proof}
First we show that \eqref{eqBg} defines a complex.  
\begin{itemize}
\item[a.] Let $\omega_k\in Z^k(M)$ be a representative of a class in $H^k(M)$.
It is obvious, from the definitions, that $\pi_* (\pi^*\omega_k) = \pi_*(\omega_k,0) = 0$.
\item[b.] Let $(\omega_k,\omega_{k-1}) \in Z_D^k(M)$ represent a class in 
$H_D^k(M)$.  We have $\delta (\pi_* (\omega_k,\omega_{k-1})  ) = F\wedge \omega_{k-1}
= - d\omega_k$, hence $\delta (\pi_* (\omega_k,\omega_{k-1})  ) \equiv0$ in $H^{k+1}(M)$.
\item[c.] Finally, let $\omega_{k-1}\in Z^{k-1}(M)$ represent a class in $H^{k-1}(M)$.
Then $\pi^*(\delta(\omega_{k-1})) = \pi^* (F \wedge \omega_{k-1}) = 
(F\wedge \omega_{k-1},0) = D(0,\omega_{k-1})$, hence 
$\pi^*(\delta(\omega_{k-1}))\equiv0$ in $H_D^{k+1}(M)$.
\end{itemize}
To show that the complex is exact, consider
\begin{itemize}
\item[1.] Let $\omega_k\in Z^k(M)$ be such that $\pi^*(\omega_k)\equiv0$ in $H^k_D(M)$,
i.e. $\pi^*(\omega_k) = (\omega_k,0) = D(\nu_{k-1},\nu_{k-2}) = 
(d\nu_{k-1} + F\wedge \nu_{k-2},-d\nu_{k-2})$
for some $(\nu_{k-1},\nu_{k-2})$.  Then $\omega_k \equiv  F\wedge \nu_{k-2}$ in 
$H^k(M)$ for some $\nu_{k-2}\in Z^{k-2}(M)$.
\item[2.] Let $(\omega_k,\omega_{k-1}) \in Z_D^k(M)$  be such that 
$\pi_*(\omega_k,\omega_{k-1})= \omega_{k-1} \equiv 0$ in $H^{k-1}(M)$, i.e.
$\omega_{k-1} = d\nu_{k-2}$ for some $\nu_{k-2}$.  Thus
$(\omega_k,\omega_{k-1}) = (\omega_k + F\wedge \nu_{k-2},0 ) - D(0,\nu_{k-2})$,
and $(\omega_k,\omega_{k-1}) \equiv (\omega_k + F\wedge \nu_{k-2},0 )$ in 
$H^k_D(M)$.  But $(\omega_k + F\wedge \nu_{k-2},0 ) = \pi^* (\omega_k + F\wedge \nu_{k-2})$.
\item[3.]  Finally, let $\omega_{k-1}\in Z^{k-1}(M)$ be such that $\delta(\omega_{k-1} )
= F\wedge \omega_{k-1} \equiv 0$ in $H^{k+1}(M)$, i.e.
$F\wedge \omega_{k-1}  = -d \nu_k$ for some $\nu_k$.  
Then $(\nu_k,\omega_{k-1}) \in Z_D^k(M)$, while $\pi_* (\nu_k,\omega_{k-1})  = 
\omega_{k-1}$.
\end{itemize}
\end{proof}

\subsection{Twisted cohomology}

Let us denote the space of even and odd forms on $E$ by 
$\Omega^{\bar 0}(E)$ and $\Omega^{\bar 1}(E)$, respectively.  I.e.\
\begin{equation} \label{eqBCa}
\Omega^{\bar\imath}(E)  = \bigoplus_{i = \bar \imath \ \text{mod}\  2} \Omega^i(E)\,.
\end{equation}
Then, given a representative $H$ for a class $[H] \in H^3(E)$, we can construct
a ``twisted differential'' $d_H : \Omega^{\bar\imath} \to \Omega^{\overline{\imath+1}}$,
by
\begin{equation} \label{eqBCc}
d_H \Omega = d\Omega + H \wedge \Omega \,.
\end{equation}
Clearly, $(d_H)^2=0$ (since $dH=0$).  The cohomology of the $\ZZ_2$-graded
complex $(\Omega^\bullet(E),d_H)$ is known as the twisted cohomology 
$H^{\bar\imath}(E,[H])$ of $E$, with
respect to the 3-form $H$.  It is easy to see that while explicit representatives for
twisted cohomology classes depend on the choice of $H$, the twisted cohomology itself 
only depends on the class $[H]$.

Let us now examine what a twisted cohomology class looks like under the dimensional 
reduction of Section 2.2.

Decomposing $H=H_3 + A\wedge H_2$, and $\Omega = \Omega^{\prime} + 
A \wedge \Omega^{\prime\prime}$, as in Eqn.\ \eqref{eqBBi}, we have
\begin{equation*}
d_H \Omega = (d\Omega^{\prime} + H_3 \wedge \Omega^{\prime} +
F\wedge \Omega^{\prime\prime}) \\ + A\wedge (-d\Omega^{\prime\prime} - H_3 \wedge \Omega^{\prime\prime}
+ H_2 \wedge \Omega^{\prime} ) \,.
\end{equation*}
Thus, the condition for $\Omega$ to be a twisted cohomology class, i.e.\ $d_H \Omega=0$,
decomposes into two equations
\begin{align}
d\Omega^{\prime} + H_3 \wedge \Omega^{\prime} +
F \wedge \Omega^{\prime\prime} & = 0\,, \nonumber \\
d\Omega^{\prime\prime} + H_3 \wedge \Omega^{\prime\prime}
- H_2 \wedge \Omega^{\prime} & = 0 \,.
\end{align}
Note that both equations do not depend on the choice of $A$, and are described
completely in terms of forms on $M$.

Now, consider the pair $((H_3,H_2),F) \in H^3_D(M) \oplus H^2(M)$.
As shown in the previous section, T-duality is the transformation 
\begin{equation}
((\widehat H_3, \widehat H_2), \widehat F)  = ((H_3,F),H_2)\,.
\end{equation}
Therefore, T-duality provides an isomorphism on twisted cohomology, which is 
explicitly given by
\begin{equation} \label{eqBCb}
(\widehat \Omega^{\prime}, \widehat\Omega^{\prime\prime}) = (\Omega^{\prime\prime},
-\Omega^{\prime}) \,.
\end{equation}
I.e.\ $d_H\Omega=0$ iff $d_{\widehat H} \widehat \Omega =0$.
Of course, Eqn.\ \eqref{eqBCb} agrees with the ``Hori formula" \cite{Hor,BEMa}
\begin{equation}
\widehat \Omega = \int_{S^1} \ e^{A\wedge\widehat A} \ \Omega \,.
\end{equation}

\section{Principal torus bundles}

In this section we will generalize the construction of the `dimensionally reduced'
Gysin sequence of the previous section to principal torus bundles.  In order 
to do this we will first have to establish how forms on the bundle space of a 
principal torus bundle are related to forms on the base space.  This is a special 
case of the so-called Chern-Weil homomorphism which holds for general 
principal $\mathsf G$-bundles.

\subsection{Dimensional reduction -- the Chern-Weil homomorphism}

Let $\sfT=\TT^n$ denote a rank-$n$ torus. Let $\ft$ denote the 
Lie algebra of $\sfT$ and denote by $\ft^*$ the dual Lie algebra.
Suppose we are given a principal $\sfT$-bundle $\pi:E\to M$.  The
action of $\sfT$ on $E$ associates to each element $X\in\ft$ a vector
field on $E$ which, by abuse of notation, we will also denote as $X$.
The Lie derivative and contraction with respect to the vector field $X$
will be denoted as $\cL_X$ and $\imath_X$, respectively.

For each cohomology class in $H^k(E)$, we may choose a 
closed representative $\Omega\in \Omega^k(E)$, such that $\cL_X \Omega =0$ for 
all $X\in \ft$.  
Locally, we can choose coordinates $x^M=(x^\mu,x^a)$
($\mu=1,\ldots,N-n$, $a=1,\ldots,n$) such that a basis of Killing vectors 
for the $\TT^n$-isometry is given by $X_a = \partial/\partial x^a$. Then $\cL_{X_a}\Omega=0$,
for all $a=1,\ldots,n$, translates into 
$\Omega(x^\mu,x^a) = \Omega(x^\mu)$.  We can decompose
$\Omega=\Omega_{M_1M_2\ldots M_k} dx^{M_1}\wedge \ldots
\wedge dx^{M_k}$ with respect to the number of directions in
$\sfT$ (`dimensional reduction'), as
$\Omega=(\Omega_{\mu_1\mu_2\ldots\mu_k}, \Omega_{\mu_1\ldots\mu_{k-1}a_1},\ldots,
\Omega_{a_1\ldots a_k})$.  
We can think of the component $\Omega_{\mu_1\ldots\mu_q a_1 \ldots a_q}$, $p+q=k$, 
as defining an element $\omega_{p,q}\in \Omega^p(M) \otimes \wedge^q \ft^*$ by 
\begin{equation} \label{eqCAa}
\omega_{p,q} = {\frac{1}{p!q!}} \Omega_{\mu_1\ldots\mu_q a_1 \ldots a_q} 
dx^{\mu^1} \wedge \ldots\wedge dx^{\mu_p} \otimes (X^{* a_1} \wedge \ldots\wedge 
X^{* a_q})\,,
\end{equation}
where $X^{*a}$, $a=1,\ldots,n$, denotes a basis of $\ft^*$.  Obviously, the original
form $\Omega\in\Omega^k(E)$ can be reconstructed locally from its components
$\omega_{p,q}\in \Omega^p(M) \otimes \wedge^q \ft^*$, $p+q=k$.

The local construction above is of course reminiscent of the K\"unneth theorem
for trivial torus bundles $E= M\times \TT^n$, in which case 
\begin{equation} \label{eqCAb}
H^k(E) \cong \bigoplus_{p+q=k}\big( H^p(M) \otimes H^q(\TT^n) \big) \cong 
\bigoplus_{p+q=k} \big( H^p(M) \otimes  \wedge^q \ft^* \big)\,.
\end{equation}

As in the case of circle bundles, the local construction above can be made global.
To this end we need a choice of (principal) connection $A\in \Omega^1(E,\ft)$ on $E$,
i.e.\ a connection $A$ satisfying 
\begin{equation} \label{eqCAc}
\imath_X A = X\,,
\end{equation}
for all $X\in \ft$ (such a connection always exists, see e.g.\ \cite{GHV}).  
Or, equivalently, if we think of a connection $A$ as defining 
a map $A : \ft^* \to \Omega^1(E)$, the principality condition can be expressed 
as 
\begin{equation}\label{eqCAd}
\imath_X A(Y^*) = \langle Y^*, X \rangle\,,
\end{equation}
for all $X\in \ft$, $Y^*\in\ft^*$.

In addition, let us introduce the following notation for
invariant, horizontal and basic forms on $E$, respectively
\begin{align} \label{eqCAe}
\Omega(E)_{\text{inv}}  & = \{ \omega \in \Omega(E)\ | \ \cL_X \omega  =0\,, \forall X\in\ft\}\,,
 \nonumber \\
\Omega(E)_{\text{hor}} & = \{ \omega \in \Omega(E)\ | \  \imath_X \omega =0\,,\forall X\in\ft\}\,,
\nonumber \\
\Omega(E)_{\text{bas}} & =  \{ \omega \in \Omega(E)\ | \ \cL_X \omega = \imath_X \omega =0\,,
\forall X\in\ft\} = \Omega(E)_{\text{hor}}  \cap \Omega(E)_{\text{inv}} \,.
\end{align}

As remarked before, we have an isomorphism $H(\Omega(E)_{\text{inv}},d) \cong H(E)$,
i.e.\ every class in $H(E)$ can be represented by a closed, invariant, form \cite{GHV}.  In addition,
basic forms are in 1--1 correspondence with forms on $M$ through the pull-back map, i.e.\
$\pi^*: \Omega(M) \to \Omega(E)_{\text{bas}}$  is an isomorphism.  

We have an isomorphism $f_A : \Omega(E)_{\text{hor}} \otimes \wedge \ft^* \to \Omega(E)$,
given by 
\begin{equation}\label{eqCAf}
f _A(\omega \otimes (X_1^*\wedge \ldots\wedge X_q^*)) = \omega \wedge
A(X_1^*) \wedge \ldots \wedge A(X_q^*) \,,
\end{equation}
which, since the group is Abelian, restricts to an isomorphism
$f_A: \Omega(E)_{\text{bas}} \otimes \wedge \ft^* \to \Omega(E)_{\text{inv}}$.
Or, since $\Omega(E)_{\text{bas}} \cong \Omega(M)$ we have 
\begin{equation}\label{eqCAg}
\bigoplus_{p+q=k} \left( \Omega^p(M) \otimes \wedge^q \ft^* \right) \cong \Omega^{k}(E)_{\text{inv}}\,.
\end{equation}

The inverse of $f_A$ is more cumbersome to write down.
First of all, given $\omega\in \Omega^k(E)_{\text{inv}}$, we can define
$\widetilde\omega_{p,q} \in \Omega^p(E) \otimes \wedge^{q} \ft^*$, with $p+q=k$, by
\begin{equation*}
\widetilde\omega_{p,q} (X_1,\ldots,X_q) = \imath_{X_1} \ldots \imath_{X_q} \omega\,,
\end{equation*}
where $X_1,\ldots,X_q\in\ft$.
Obviously, $\cL_X\widetilde\omega_p=0$, thus 
$\widetilde\omega_{p,q} \in \Omega^p(E)_{\text{inv}} \otimes \wedge^{q} \ft^*$. However,
\begin{equation*}
\imath_X \widetilde\omega_{p,q} (X_1,\ldots,X_q) = \widetilde\omega_{p-1,q+1} (X,X_1,\ldots,X_q)\,.
\end{equation*}
Thus, upon defining, $\omega_{p,q} \in \Omega^p(E)\otimes \wedge^{q} \ft^*$ by
\begin{equation}\label{eqCAh}
\omega_{p,q} (X_1,\ldots,X_q) = \sum_{r=0}^p \frac{1}{r!} (-1)^{rp - \frac12 r(r+1)}\ 
\widetilde\omega_{p-r,q+r} (\underbrace{A,\ldots,A}_r,X_1,\ldots,X_q) \,,
\end{equation}
we have $\omega_{p,q} \in \Omega^p(E)_{\text{bas}} \otimes \wedge^{q} \ft^*$, where
we have used 
\begin{multline*}
\imath_X \widetilde \omega_{p-r,q+r}(\underbrace{A,\ldots,A}_r,X_1,\ldots,X_q)  = 
\widetilde\omega_{p-r-1,q+r+1}(X,\underbrace{A,\ldots,A}_r,X_1,\ldots,X_q) \\ + (-1)^{p-r} r\ 
\widetilde \omega_{p-r,q+r}(X,\underbrace{A,\ldots,A}_{r-1},X_1,\ldots,X_q)\,.
\end{multline*}
Hence,
$f_A^{-1} : \Omega^k(E)_{\text{inv}} \to \bigoplus_{p+q=k}
(\Omega^p(M) \otimes \wedge^q \ft^*)$, is given by 
\begin{equation}
\omega \mapsto (\omega_{k,0},\omega_{k-1,1},\ldots,\omega_{0,k})\,,
\end{equation}
with $\omega_{p,q}\in \Omega^p(M) \otimes \wedge^{q} \, \ft^*$, and 
$p+q=k$. 
We will often simplify the notation and simply write 
\begin{equation}\label{eqCAi}
\omega \mapsto (\omega_k,\omega_{k-1}, \ldots, \omega_0)\,,
\end{equation}
with $\omega_p\in \Omega^p(M) \otimes \wedge^{k-p} \, \ft^*$.

We think of the curvature $F\in Z^2(M,\ft)\cong Z^2(M) \otimes \ft$,
and thus as defining a map $F : \ft^* \to \Omega^2(M)$, satisfying $dF(X^*)=0$ for all
$X^*\in \ft^*$.
The differential on $\Omega^p(M) \otimes \wedge^q \ft^*$ is then defined by
\begin{multline}\label{eqCAj}
D \left(\omega \otimes (X_1^*\wedge \ldots \wedge X_q^*) \right) = 
d\omega \otimes (X_1^*\wedge \ldots \wedge X_q^*)  \\ + (-1)^p
\sum_{i=1}^q (-1)^{i-1}\,  (\omega \wedge F(X_i^*)) \otimes (X_1^* \wedge \ldots \wedge
\widehat{X_i^*} \wedge \ldots \wedge X_q^*) \,.
\end{multline}
It is easily verified that $D^2=0$ using the fact that
$F$ is closed and the symmetry of $F(X_i^*) \wedge F(X_j^*)$ in $i$ and $j$.

Next we show that $f_A\circ D = d \circ f_A$.  On the one hand we have,
for $\omega \otimes (X_1^*\wedge \ldots \wedge X_q^*) \in \Omega^p(M) \otimes \wedge^q \ft^*$
\begin{multline*}
(d\circ f_A) (\omega \otimes (X_1^*\wedge \ldots \wedge X_q^*)) = 
d( \omega \wedge A(X_1^*) \wedge \ldots \wedge A(X_q^*) ) \\ = 
d\omega \wedge A(X_1^*) \wedge \ldots \wedge A(X_q^*) + 
(-1)^p \sum_{i=1}^q (-1)^{i-1}  \omega \wedge A(X_1^*) \wedge 
\ldots \wedge F(X_i^*)\wedge  \ldots \wedge A(X_q^*)\,,
\end{multline*}
while
\begin{multline*}
(f_A\circ D)(\omega \otimes (X_1^*\wedge \ldots \wedge X_q^*)) =
f_A\big( d\omega \otimes (X_1^*\wedge \ldots X_q^*)\\ + (-1)^p \sum_{i=1}^q (-1)^{i-1}  (\omega
\wedge F(X_i^*) ) \otimes (X_1^*\wedge \ldots 
\wedge \widehat{X_i^*} \wedge \ldots \wedge X_q^*)\big) \\
= d\omega \wedge A(X_1^*) \wedge \ldots \wedge A(X_q^*) + 
(-1)^p \sum_{i=1}^q (-1)^{i-1}  (\omega \wedge F(X_i^*)) \wedge A(X_1^*) \wedge 
\ldots \wedge \widehat{A(X_i^*)} \wedge  \ldots \wedge A(X_q^*)\,.
\end{multline*}
Obviously, the two expressions are equal, and 
thus we have a commutative diagram
\begin{equation*} \begin{CD}
\bigoplus_{p+q=k} \big(\Omega^p(M)\otimes \wedge^q 
\ft^*\big) @>\cong>f_A>  \Omega^k(E)_{\text{inv}} \\
@VDVV  @VVdV \\
\bigoplus_{p+q=k+1} \big(\Omega^{p}(M)
\otimes \wedge^q \ft^*\big)@>\cong>f_A> 
\Omega^{k+1}(E)_{\text{inv}} \end{CD}
\end{equation*}
To summarize, if we denote 
\begin{equation}
\Omega^k(M,\ft^*) \cong \bigoplus_{p+q=k} \big( \Omega^p(M) \otimes \wedge^q\, \ft^* \big)\,,
\end{equation}
we have a complex $(\Omega(M,\ft^*),D)$ whose cohomology, $H_D^k(M,\ft^*)$, is 
isomorphic to the cohomology $H^k(E)$.\footnote{We denote this cohomology by
$H^k_D(M,\ft^*)$ to distinguish it from the cohomology $H^k(M,\ft^*)$ of $k$-forms
with coefficients in $\ft^*$.}

\begin{remark}
It can be shown that while the Chern-Weil homomorphism depends on the choice of 
connection, the isomorphism on cohomology does not (see, e.g., \cite{GHV}).
\end{remark}

\begin{remark}
As mentioned before, the construction we have described here is merely a special 
case of a theory constructed by Chevalley, Chern, Weil, Cartan and others 
(see, in particular, \cite{Cara, Carb, GHV, GS}).  Given a Lie group $\sfG$, its cohomology
can be calculated as $H(\sfG) \cong \wedge P(\fg)$, where $P(\fg)$ is a set
of primitive elements in the symmetric algebra $S\fg^*$, and $\fg$ is the Lie algebra
of $\sfG$.  Now, given a principal $\sfG$-bundle $E\to M$, there exists an 
isomorphism between the cohomology $H(E)$ and the cohomology of a 
complex $\Omega(M) \otimes \wedge P(\fg)$ with Koszul differential $D$.
The case of abelian $\sfG$, in this paper, is special since in that case 
the Chern-Weil homomorphism $\Omega(M) \otimes \wedge P(\fg) \to \Omega(E)_{\text{inv}}$
is actually an isomorphism, unlike in the more general case.
\end{remark}

\subsection{Dimensionally reduced Gysin sequences} 

In order to write down a Gysin sequence for principal $\sfT$-bundles $\pi : E\to M$,
part of which is the pull-back map $\pi^* : H^k(M) \to H^k(E) \cong H_D^k(M,\ft^*)$
given simply by $\pi^* \omega = \omega \in \Omega^k(M) \otimes \wedge^0 \ft^*$,
we need to define a map $\pi_*$ on $H_D^k(M,\ft^*)$ by `stripping off' the component
in $\Omega^k(M)\otimes \wedge^0 \ft^*$.  Thus, in addition to
\begin{equation}
\Omega^k(M,\ft^*) \cong \bigoplus_{p+q=k}  \big(\Omega^p(M) \otimes \wedge^q\, \ft^*\big) \,,
\end{equation}
let us define the following truncated versions of this space (for $0\leq r\leq s \leq n$)
\begin{equation} \label{eqCBb}
\Omega^{k,(r,s)}(M,\ft^*) \cong \bigoplus_{i=r}^s \big(\Omega^{k-i}(M) \otimes \wedge^i\, \ft^*\big)\,,
\end{equation}
such that $\Omega^k(M,\ft^*) \cong \Omega^{k,(0,n)}(M,\ft^*)$, and
$\Omega^{k}(M)\cong \Omega^{k,(0,0)}(M,\ft^*)$.   Also denote the basic 
truncation as 
\begin{equation}
\overline \Omega^k(M,\ft^*) \equiv \Omega^{k,(1,n)}(M,\ft^*)\,.
\end{equation}
The differential $D$ of Eqn.\ \eqref{eqCAj} restricts to a 
differential on $\Omega^{\bullet,(r,s)}(M,\ft^*)$, and 
defines a cohomology $H_D^{k,(r,s)}(M,\ft^*)$.  In particular, we have
$H^k(M)\cong H_D^{k,(0,0)}(M,\ft^*)$, and $H_D^k(M,\ft^*) \cong H_D^{k,(0,n)}(M,\ft^*)$. 
We also define\footnote{Note the shift in degree, chosen such that $\overline H_D^{k-1}(M,\ft^*)
\cong H^{k-1}(M)$ for principal circle bundles.}
\begin{equation}
\overline H_D^{k-1}(M,\ft^*) \equiv H_D^{k,(1,n)}(M,\ft^*) \,.
\end{equation}
We then have
\label{thCBa}
\begin{theorem} 
We have the following long exact sequence for cohomologies related to
a principal torus bundle $\pi:E\to M$.
\begin{equation} \label{eqCBg}
\begin{CD}
@>>> H^k(M) @>\pi^*>> H_D^k(M,\ft^*)  @>\pi_*>> \overline H_D^{k-1}(M,\ft^*) @>\delta>>
H^{k+1}(M) @>>> 
\end{CD}
\end{equation}
where the maps are given, on representatives, by
\begin{align*}
\pi^* : & H^{k}(M) \to H_D^{k}(M,\ft^*)\,, & &
\pi^* (\omega_{k})  = (\omega_{k},0,\ldots,0)\,, \\
\pi_* : & H_D^{k}(M,\ft^*)  \to \overline H_D^{k-1}(M,\ft^*)\,,  & &
\pi_*(\omega_{k},\ldots,\omega_{0}) = (\omega_{k-1},\ldots,\omega_{0})\,, \\
\delta : & \overline H_D^{k-1}(M,\ft^*) \to H^{k+1}(M)\,, & &
\delta(\omega_{k-1},\ldots,\omega_{0}) = F(X^*)\wedge\widetilde\omega_{k-1}\,, && 
(\text{\rm if}\ \omega_{k-1} \equiv \widetilde\omega_{k-1}\otimes X^*)\,.
\end{align*}
\end{theorem}

\begin{proof}
The proof is  exactly analogous to the proof in Section \ref{secBC} 
for the circle bundle case.
The proof that \eqref{eqCBg} defines a complex is straightforward. 
The hardest part in the proof of exactness is at $\overline H_D^{k-1}(M,\ft^*)$.
Thereto, suppose ${\mathbf \omega} = (\omega_{k-1},\ldots,\omega_{0})$, with
$\omega_{k-1}=\widetilde\omega_{k-1}\otimes X^*$, 
is a representative
of a class in $\overline H_D^{k-1}(M,\ft^*)$ such that $\delta{\bfom}=0$ in 
$H^{k+1}(M)$, i.e.\ we have $F(X^*)\wedge \widetilde\omega_{k-1} = -d\nu$ for
some $\nu\in \Omega^{k}(M)$.
Then ${\mathbf \nu} = (\nu,\omega_{k-1},\ldots,\omega_{0}) \in 
\Omega^{k}(M,\ft^*)$ satisfies $D{\bfnu} = 0$, while $\pi_*\bfnu = \bfom$.
\end{proof}

In fact, Theorem \ref{thCBa} is easily generalized by considering different truncations 
as in \eqref{eqCBb}.  Namely
\label{thCBb}
\begin{theorem} For $0\leq r < s \leq n$, we have exact sequences
\begin{equation} \label{eqCa}  \begin{CD}
 H_D^{k,(r,r)}(M,\ft^*) @>\pi^*>> H_D^{k,(r,s)}(M,\ft^*) @>\pi_*>> H_D^{k,(r+1,s)}(M,\ft^*) 
@>\delta>> H_D^{k+1,(r,r)}(M,\ft^*) @>>>
\end{CD}
\end{equation}
where 
\begin{align*}
\pi^* : & H_D^{k,(r,r)}(M,\ft^*) \to H_D^{k,(r,s)}(M,\ft^*) \,, & &
\pi^* (\omega_{k-r})  = (\omega_{k-r},0,\ldots,0) \,, \\
\pi_* : & H_D^{k,(r,s)}(M,\ft^*)  \to H_D^{k,(r+1,s)}(M,\ft^*) \,,  & &
\pi_*(\omega_{k-r},\ldots,\omega_{k-s}) = (\omega_{k-r-1},\ldots,\omega_{k-s})\,, \\
\delta : & H_D^{k,(r+1,s)}(M,\ft^*) \to H_D^{(k+1,(r,r)}(M,\ft^*) \,, & & 
\end{align*}
\begin{align*}
\delta(\omega_{k-r-1},\ldots,\omega_{k-s}) & = \sum_i (-1)^{i-1} F(X_i^*)\wedge\widetilde 
\omega_{k-r-1} \otimes (X_1^*\wedge \ldots \wedge \widehat{X_i^*}\wedge
\ldots\wedge X_{r+1}^*) \,,\\
& \text{\rm if} \quad  \omega_{k-r-1} = \widetilde \omega_{k-r-1} \otimes (X_1^*\wedge \ldots \wedge X_{r+1}^*) \,.
\end{align*}
\end{theorem}

The basic Gysin sequence for $\pi:E\to M$ is the one given by $r=0,s=n$ in
Theorem \ref{thCBa}.  The other sequences can be viewed as 
truncated Gysin sequences (from the left if $r>0$, and from the right if $s<n$.
In particular, we note that for $r=0,s=1$ we get the Gysin sequence 
of \cite{PRW}, which was discussed in the de-Rham framework in \cite{BHMa}.
The image of $i : H^{3,(0,1)}(M,\ft^*) \hookrightarrow H^3(E)$ by means of 
the Chern-Weil homomorphism $f_A$
(Eqn.\ \eqref{eqCAf}) is what was called a T-dualizable H-flux in \cite{BHMa}.

\section{Application to T-duality}

\subsection{T-duality for principal torus bundles}
\label{secDA}

Suppose we are given a principal torus bundle $\pi: E\to M$, with curvature class $[F]\in
H^2(M)$, and with H-flux $[H] \in H^3(E)$.  
We will think of this as specifying
an element $([H],[F])\in H^3(E) \oplus H^2(M)$.  We will choose a representative $(H,F)$
which, upon dimensional reduction, can be viewed as a  
tuple $((H_3,H_2,H_1,H_0),(F_2,0,0))$, with $H_i\in\Omega^i(M) \otimes \wedge^{3-i}\,\ft^*$,
$F_i\in \Omega^i(M)\otimes \wedge^{2-i} \,\ft$, both closed under $D$.

The image of $H\equiv (H_3,H_2,H_1,H_0)$, in the Gysin sequence \eqref{eqCBg}, is given
by $\widehat F \equiv (\widehat F_2, \widehat F_1,\widehat F_0) = (H_2,H_1,H_0)$.  This
3-tuple is supposed to classify (up to torsion) our T-dual object.   Subsequently, one would
expect the T-dual H-flux carried by this object to be given by the 4-tuple 
$\widehat H \equiv (\widehat H_3, \widehat H_2, 
\widehat H_1, \widehat H_0) = (H_3, F_2, F_1,F_0) = (H_3,F_2,0,0)$.

In the case $H_1=H_0=0$, our T-dual object is characterized by $\widehat F = (H_2,0,0)$,
where $H_2 \in \Omega^2(M)\otimes \ft^*$, and hence can be identified with the 
curvature of a principal $\widehat\sfT$-bundle $\hat\pi: \widehat E \to M$.  
This case was analyzed in detail in \cite{BHMa}, and the corresponding H-fluxes were
dubbed `T-dualizable'. 

As soon as $H_1\neq0$ or $H_0\neq0$, the T-dual object takes us outside
the realm of principal torus bundles.   The case $H_1\neq0$, $H_0=0$, was analyzed 
in detail in \cite{MRa,MRb}.  In this case the T-dual object was argued to be a 
continuous field of noncommutative tori over the base space $M$.  Concretely, 
in this case $H_1$ determines an integral class $[H_1]\in H^1(M,\wedge^2 \ft^*)$ or,
equivalently, a homotopy class in $[M,\wedge^2 \sfT^*]$.  If $f: M\to\wedge^2 
\widehat \sfT$ is a representative for this class, then the fibre over a point $z$ in the base $M$
is given by the noncommutative torus $A_{f(z)}$.\footnote{Locally, one can think of the
commutativity parameter of the torus as given by the components of the
B-field  in the torus directions.}
A global description of this field of noncommutative tori is given in terms of
a crossed product algebra $\fA \rtimes \RR^n$, such that $\text{spec}(\fA) = E$ 
(see Appendix \ref{appA} for some details).
The results of this paper suggest that appropriate equivalence classes
of these objects are classified by a triple $(F_2,F_1,0)$, $F_i\in \Omega^i(M)\otimes \wedge^{2-i}
\ft$, closed under $D$.  It would be extremely interesting to establish this directly, and to find a more
`geometric' description of the T-dual object even in this case.

In the most general case, $H_0\neq0$, the T-dual object carries information about an 
integral class $[H_0] \in H^0(M,\wedge^3\ft^*)$, i.e.\ a locally constant function with 
values in $\wedge^3 \sfT^*$. It is well-known that such classes often correspond to
nonassociative structures (cf.\ \cite{Jac, Car}).  In \cite{BHMc} we construct a $C^*$-algebra $\fA$,
with Dixmier-Douady invariant $H=(H_3,H_2,H_1,H_0)$ and $\text{spec}(\fA)=E$,
as a twisted induced algebra,
that carries a twisted action of $\RR^n$.  This generalizes a construction in \cite{MRa,MRb},
which applies to $H=(H_3,H_2,H_1,0)$, by introducing a twisting $u$ 
given by the $H_0$-component.\footnote{Note that while the analysis in this paper
is carried out in de-Rham cohomology, the component $H_0$ does not carry torsion,
it can in fact simple be described as the pull-back of $H$ from $H^3(E,\ZZ)$ to
$H^3(\sfT,\ZZ)$ under fibre inclusion $i:\sfT \hookrightarrow E$.  Hence, the 
construction of \cite{BHMc}, by adding an $H_0$ part, is valid for arbitrary 
integer H-fluxes in $H^3(E,\ZZ)$.}
We argue that the T-dual is given by the twisted crossed product $\fA\rtimes_u \RR^n$,
and that this twisted crossed product can be interpreted as a continuous field of
noncommutative, nonassociative tori over the base $M$.
It would be interesting to establish directly precisely which objects are classified
by a 3-tuple $(F_2,F_1,F_0)$, $F_i\in \Omega^i(M)\otimes \wedge^{2-i}
\ft$, closed under $D$, and in which sense the T-dual H-flux can be interpreted as 
a flux on that object.

\subsection{The T-duality group}

The T-duality group is $O(n,n; \ZZ)$.  It turns out that the action of the T-duality 
group on the tuples 
\begin{equation*}
((H_3, H_2, H_1, H_0), (F_2, F_1, F_0)) \,,
\end{equation*}
where $H_i \in \Omega^i(M, \wedge^{3 - i} \ft^*)$ for 
$i = 0, 1, 2, 3$ and 
 $F_i \in \Omega^i(M, \wedge^{2- i} \ft)$  for 
$i = 0, 1, 2$, has a very simple expression.
 
Given $g \in O(n,n; \ZZ)$, it acts naturally on $\ft^* \oplus \ft$, preserving the
natural quadratic form.  It induces an action on $\wedge^i \ft^* \oplus \wedge^i \ft$.  
Let us denote by $g_i = \wedge^{3 - i} g$ the action of $O(n,n; \ZZ)$ on
$\wedge^{3 - i} \ft^* \oplus  \wedge^{3 - i} \ft$  (for $i = 0, 1, 2$).
We have a corresponding induced action on $ \Omega^i(M, \wedge^{3 - i} \ft^*) \oplus  
\Omega^i(M, \wedge^{3 - i} \ft)$  for $i = 0, 1, 2$,
which we also denote by $g_i = \begin{pmatrix} A_i & B_i\\C_i& D_i\end{pmatrix}$. 
 
The action of  $g \in O( n,n; \ZZ)$ on 
$((H_3, H_2, H_1, H_0), (F_2, F_1, F_0))$
is then explicitly explicitly by 
\begin{multline}
g\cdot ((H_3, H_2, H_1, H_0), (F_2, F_1, F_0)) \cong 
\left( H_3, g_2\cdot \begin{pmatrix} H_2 \\ F_2 \end{pmatrix}, 
g_1\cdot \begin{pmatrix} H_1 \\ F_1\end{pmatrix},
g_0 \cdot\begin{pmatrix} H_0 \\ F_0 \end{pmatrix}\right) \cong \\
((H_3, A_2 H_2 + B_2 F_2, A_1 H_1 + B_1 F_1, A_0 H_0 + B_0 F_0), (C_2 H_2 + D_2 F_2, 
C_1 H_1 + D_1 F_1, C_0 H_0 + D_0 F_0)) \,.
\end{multline}
Note that the action of the T-duality group resembles fractional linear transformations.

In particular, if we start out with a principal torus bundle with H-flux, then the action of 
$g  \in O(n,n; \ZZ)$ on $((H_3, H_2, H_1, H_0), (F_2, 0, 0))$
is given explicitly by 
\begin{equation}
((H_3, A_2 H_2 + B_2 F_2, A_1 H_1, A_0 H_0 ), (C_2 H_2 + D_2 F_2, 
C_1 H_1, C_0 H_0)) \,.
\end{equation}

The T-duality transformation discussed in Section \ref{secDA}, corresponds to
the element $g \in O(n,n,\ZZ)$ given by 
\begin{equation}
g = \begin{pmatrix} 0 & \mathbf 1_n \\ \mathbf 1_n & 0 \end{pmatrix} \,.
\end{equation}

\subsection{Example}

The simplest example of the various cases of T-duality is given by the three
torus $\TT^3$, with H-flux $k dV = k dx\wedge dy\wedge dz$, which can be considered
as a principal torus bundle over a (strictly) lower dimensional torus in three different ways

\begin{enumerate}
\item $(\TT^3, k\; dx \wedge dy \wedge dz)$ considered as a trivial circle bundle over 
$\TT^2$. The T-dual of  $(\TT^3, k\; dx \wedge dy \wedge dz)$ is the nilmanifold
$(H_\RR/H_\ZZ, 0)$, where $H_\RR$ is the 3 dimensional Heisenberg group and
$H_\ZZ$  the lattice in it defined by
 \begin{equation}
H_{\mathbb Z} = \left\{   \begin{pmatrix} 1 & x & \frac{1}{k}z \\
0 & 1 & y\\
0 & 0 &       1
\end{pmatrix} : x, y, z \in \mathbb Z \right\}.
\end{equation}

\item $(\TT^3, k\; dx \wedge dy \wedge dz)$ considered as a trivial $\TT^2$-bundle over 
$\TT$. The T-dual of  $(\TT^3, k\; dx \wedge dy \wedge dz)$ is a continuous field 
of stabilized noncommutative tori, $C^*(H_\ZZ) \otimes \mathcal K$,
since 
\begin{equation*}
H_1 \sim \int_{\TT^2=\{(y,z)\}}  k\; dx \wedge dy \wedge dz = k\; dx \neq0 \,.
\end{equation*}

\item  $(\TT^3, k\; dx \wedge dy \wedge dz)$ considered as a trivial $\TT^3$-bundle over 
a point. The T-dual of  $(\TT^3, k\; dx \wedge dy \wedge dz)$ is a nonassociative
 torus, $A_\phi$, where $\phi$ is the tricharacter associated to 
$ k\; dx \wedge dy \wedge dz $, since
\begin{equation*}
H_0 \sim \int_{\TT^3}  k\; dx \wedge dy \wedge dz = k \neq 0 \,.
\end{equation*}
\end{enumerate}
 
Other examples, treated in previous papers, such as the nilmanifold \cite{BEMa}
and the group manifold, viewed as a principal torus bundle over the flag 
manifold \cite{BHMa}, both with H-flux, can be re-interpreted similarly.

\section{Conclusions and further generalizations}
 
In this paper we have shown how Gysin sequences encode the global 
properties of T-duality, building on previous work \cite{BEMa, BEMb, BHMa}.
We have constructed a Gysin sequence for principal torus bundles, using 
what is known in physics as dimensional reduction, and in mathematics as 
a special case of the Chern-Weil homomorphism, and have used it to determine
the T-dual of a principal torus bundle with arbitrary H-flux, generalizing the special
cases considered in \cite{BEMa, BEMb, BHMa, MRa, MRb}.  The algebraic structures
arising in the T-dual have been discussed in a separate paper \cite{BHMc}.
Gysin sequences are useful in a more general context as well, e.g.\ in cases 
where we do not have a principal torus bundle, but only an infinitesimally free torus action
(e.g. Seifert fibered spaces, in which case the base manifold is a 2D orbifold)

 Once one realizes that the T-dual of a principal torus bundle, characterized by a 
curvature $F=(F_2,0,0)$, $F_2\in \Omega^2(M) \otimes \ft$, with background H-flux $H=(H_3,H_2,H_1,H_0)$,
$H_i\in \Omega^i(M) \otimes \wedge^{3-i}\ft^*$, is an object characterized by the 
3-tuple $(H_2,H_1,H_0)$, it seems natural to somehow study a Gysin sequence 
related to such an object and to interpret the dual tuple $\widehat H= (H_3,F_2,0,0)$ 
as an H-flux  on this dual object by reversing the argument in the Gysin sequence.
This we leave as a problem for further investigation.  From the physics point of view,
the results of this paper show that if one is trying to build a manifestly T-duality 
invariant description of M-theory, one is forced to include not only noncommutative
structures, but also nonassociative structures into the game.

Another obvious extension of this work is to incorporate torsion by generalizing the
analysis of this paper from de-Rham cohomology to integer, i.e.\ \v Cech cohomology.
Although there are some subtleties, we believe this is possible and we hope to
come back to this in a future publication (cf.\ \cite{Fra, Frb} for some relevant 
results in this direction).  One could even wonder whether a similar dimensional 
reduction might be possible at the level of K-theory.

Maybe one of the most striking results of this paper is that we are led to T-dual objects
which appear to be quite `pathological' -- noncommutative or even nonassociative -- 
but nevertheless are characterized by forms on the base manifold, one realizes that 
this approach might be useful is getting a hold on even more exotic structures such
as torus fibrations (manifolds with non free torus actions), as long as the base manifold 
has a well-understood theory of differential forms and de-Rham cohomology.

\section*{Acknowledgements}

PB would like to thank the Erwin Schr\"odinger Institute in Vienna, the
CIRM in Luminy and the Aspen Center for Physics 
for hospitality and financial support during various stages of this project.
KCH would like to thank the University of Adelaide for hospitality.
PB and VM were financially supported by the Australian Research Council.


\begin{appendix}

\section{Duality from the operator algebraic perspective}
\label{appA}

In this paper, as well as previous ones \cite{BEMa, BHMa}, we have attempted to
sketch the geometric counterpart, as well as applications to T-duality, of results 
etablished in the context of, in particular, operator algebras, derived in a series of 
beautiful papers (see, in particular, \cite{RR, RWa, RWb, RWc, PRW, PR, CKRW, MRa, MRb, BHMc}).  
For the 
benefit of readers wishing to familiarize themselves with the algebraic perspective,
we include this appendix.  

We begin with a reformulation of \cite{BEMa, BEMb}. Given a circle bundle 
$S^1 \hookrightarrow E\to M$ over $M$, and a closed, integral
3-form $H$ on $E$, then there is a unique algebra bundle
$\cE \to E$ with fibre equal to the algebra of compact operators 
$\cK$ and Dixmier-Douady invariant equal to $[H] \in H^3(E, \ZZ)$.
Then the space of continuous sections $\;\frak A = C(E, \cE)$ is a 
stable, continuous trace $C^*$-algebra
with spectrum ${\rm spec}(\fA)= E$. The $\RR$ action on $E$
lifts uniquely to an $\RR$ action on $\fA$ (cf. \cite{RR}), 
and one has a commutative diagram,
\begin{equation}\label{eqn:CD} 
\xymatrix @=9pc @ur { {\rm spec}(\fA) \ar[d]_{\pi} & 
{\rm spec}(\fA \rtimes \ZZ) \ar[d]_{\hat p} \ar[l]^{p} \\ {\rm spec}(\fA)/\RR & {\rm spec}(\fA \rtimes \RR)\ar[l]^{\hat \pi}}
\end{equation}

That is, $\fA \rtimes \ZZ$ and $\fA \rtimes \RR$ are also continuous trace 
$C^*$-algebras with ${\rm spec}(\fA \rtimes \RR) = \widehat E$ a circle bundle
over $M = {\rm spec}(\fA)/\RR $, such that $c_1(\widehat E) = \pi_*[H]$ and the Dixmier-Douady 
invariant of $\fA \rtimes \RR$ is $[\hat H] \in H^3(\widehat E, \ZZ)$,
such that $c_1(E) = \hat\pi_*[\widehat H]$, and ${\rm spec}(\fA \rtimes \ZZ) = 
E\times_M \widehat E$ is the correspondence space. Now the T-dual of the 
continuous trace $C^*$-algebra $\fA \rtimes \RR$ is the crossed product 
$(\fA \rtimes \RR) \rtimes \widehat \RR$, which by Takai duality is Morita equivalent to $\fA$,
and in particular, $ {\rm spec}((\fA \rtimes \RR) \rtimes \widehat \RR) = {\rm spec}(\fA)$.
That is, applying T-duality twice gets us back to where we started off.
As a result, we also get the horizontal isomorphisms 
(Connes-Thom isomorphisms in K-theory and
in cyclic homology) and the commutativity of the diagram,
\begin{equation} \label{eqn:RR}
\begin{CD}
K_\bullet(\fA)  @>T_!>\cong> K_{\bullet}({\fA}\rtimes \RR)  \\
        @V{Ch_H}VV          @VV{Ch} V     \\
HP_\bullet  ({\fA}^\infty)    @>T_*>\cong>  HP_{\bullet}
({\fA}^\infty \rtimes \RR)\\
\end{CD}
\end{equation}
where ${\fA}^\infty = C^\infty(E, \cE^\infty)$ is a smooth subalgebra of $\fA$. This motivates 
the definition of the T-dual of a principal torus bundle with H-flux, when the T-dual is not classical,
viz.\ when the T-dual is not another principal torus bundle with H-flux. 

In dealing with higher rank torus bundles, we will use the notation
$\sfG=\RR^n$, $\sfN=\ZZ^n$, and $\sfT=\sfG/\sfN = \TT^n$.
Now, given a torus bundle 
$\sfT \hookrightarrow E\to M$ over $M$, and a closed, integral
3-form $H$ on $E$, then there is a unique algebra bundle
$\cE \to E$ with fibre equal to the algebra of compact operators 
$\cK$ and Dixmier-Douady invariant equal to $[H] \in H^3(E, \ZZ)$.
Let $\fA$ denote the space of all continuous sections of $\cE$,
which is a stable, continuous trace $C^*$-algebra with spectrum
$\text{spec}(\fA)= E$.
Now if (and only if) the restriction of $[H]$ to each fibre is trivial (i.e. $H_0=0$), then
the $\sfG$-action on $E$ lifts to an $\sfG$-action on the 
total space $\cE$, i.e. there is an induced $\sfG$-action on $\fA$.
The lift of the $\sfG$-action on $E$
to $\fA$ is not unique, cf. \cite{MRa, MRb}, but this does not affect the K-theory.
The T-dual of the torus bundle with H-flux, $(E, H)$, is then defined as
$\fA \rtimes \sfG$. Note that this is in general not a stable, continuous trace
$C^*$-algebra.
Then we have an analogous diagram for this situation as in \eqref{eqn:RR}. More
importantly, by Takai duality, $(\fA \rtimes \sfG) \rtimes \widehat \sfG$ 
is Morita equivalent to $\fA$,
and in particular, $ {\rm spec}((\fA \rtimes \sfG )\rtimes \widehat \sfG) = {\rm spec}(\fA)$.
That is, applying T-duality twice gets us again back to where we started off, as was established in 
\cite{MRa, MRb}.

Given a torus bundle $\sfT \hookrightarrow E\to M$ over $M$, and a closed, integral
3-form $H$ on $E$, let $\cE \to E$ be the unique algebra bundle
with fibre equal to the algebra of compact operators 
$\cK$ and Dixmier-Douady invariant equal to $[H] \in H^3(E, \bZZ)$.
Let $\fA$ denote the space of all continuous sections of $\cE$.
Now if  the restriction of $[H]$ to each fibre is {\em not} trivial (i.e. $H_0 \ne 0$), then
the $\sfG$-action on $E$ lifts to a twisted $\sfG$-action on $\cE$
and hence on $\fA$, \cite{BHMc}. Again this lifted action is not unique. 
The T-dual of the torus bundle with H-flux, $(E, H)$, is defined as the
twisted crossed product $\fA \rtimes_u \sfG$.
The twisted crossed 
product $\fA \rtimes_u \sfG$ is in general a nonassociative, noncommutative
algebra, which is a field of nonassociative tori on the base. By twisted Takai duality 
\cite{BHMc},  $(\fA \rtimes_u \sfG) \rtimes_{\hat u} \widehat \sfG$ is 
Morita equivalent to $\fA$,
and in particular, $ {\rm spec}((\fA \rtimes_u \sfG) \rtimes_{\hat u} 
\widehat \sfG) = {\rm spec}(\fA)$.
That is, applying T-duality twice gets us again back to where we started off, as was established in 
\cite{BHMc}.

\end{appendix}


\end{document}